\documentclass[12pt,a4paper]{article}
\usepackage{a4wide}
\usepackage{latexsym}
\usepackage{epsf}
\usepackage{amssymb}
\makeatletter
\@addtoreset{equation}{section}
\makeatother

\begin{document}

\begin{flushright}
\small
IFT-UAM/CSIC-00-25\\
{\bf hep-th/0012051}\\
December $7$th, $2000$
\normalsize
\end{flushright}

\begin{center}


\vspace{.7cm}

{\LARGE {\bf 7-Branes and Higher Kaluza-Klein Branes}}

\vspace{.7cm}


{\bf\large Ernesto Lozano-Tellechea}${}^{\spadesuit}$
\footnote{E-mail: {\tt Ernesto.Lozano@uam.es}},
{\bf\large and Tom\'as Ort\'{\i}n}${}^{\spadesuit\clubsuit}$
\footnote{E-mail: {\tt tomas@leonidas.imaff.csic.es}}
\vskip 0.4truecm

\vskip 0.2cm

${}^{\spadesuit}$\ {\it Instituto de F\'{\i}sica Te\'orica, C-XVI,
Universidad Aut\'onoma de Madrid \\
E-28049-Madrid, Spain}

\vskip 0.2cm

${}^{\clubsuit}$\ {\it I.M.A.F.F., C.S.I.C., Calle de Serrano 113 bis\\ 
E-28006-Madrid, Spain}

\vspace{.7cm}


{\bf Abstract}

\end{center}

\begin{quotation}

\small

We present and study a new chain of 10-dimensional T~duality related
solutions and their 11-dimensional parents whose existence had been
predicted in the literature based in U~duality requirements in 4
dimensions. The first link in this chain is the S~dual of the
D7-brane. The next link has 6 spatial worldvolume dimensions, it is
charged w.r.t.~the RR 7-form but depends only on 2 transverse
dimensions since the third has to be compactified in a circle and is
isometric and hence is similar in this respect to the KK monopole. The
next link has 5 spatial worldvolume dimensions, it is charged
w.r.t.~the RR 6-form but, again, depends only on 2 transverse
dimensions since the third and fourth have to be compactified in
circles and are isometric  and so on for the following links.

All these solutions are identical when reduced over the $p$ spatial
worldvolume dimensions and preserve a half on the available
supersymmetries. Their masses depend on the square of the radii of the
isometric directions, just as it happens for the KK monopole. We give
a general map of these branes and their duality relations and show how
they must appear in the supersymmetry algebra.

\end{quotation}

\newpage

\pagestyle{plain}


\section*{Introduction}

In the last few years there has been a lot of interest in discovering
classical solutions of effective superstring theories (supergravity
theories) with such properties that one could argue that they
represent the fields produced by solitonic objects present in the
superstring spectrum. The interplay between the knowledge of the
superstring spectrum and the knowledge of classical solutions has been
very fruitful since each of them has contributed to the increase of
the other. The two most important tools used in this field have been
supersymmetry and duality. Unbroken supersymmetry ensures in many
cases the absence of corrections of the classical solutions and the
lack of quantum corrections to the mass of the corresponding objects
in the string theory spectrum. Hence, more effort has been put in
finding supersymmetric (i.e.~admitting Killing spinors) solutions,
associated to BPS string states.  Duality transformations preserve in
general supersymmetry, relating different states in dual theories. In
general \cite{kn:HT}, but not always \cite{kn:BHO} duality relations
between different higher-dimensional theories manifest themselves as
non-compact global symmetries of the compactified supergravity theory
that leave invariant its equations of motion so one can use them to
transform known solutions into new solutions, preserving their
supersymmetry properties.

Thus, it so happens that most classical solutions of superstring
effective field theories belong to chains or families of solutions
related by duality transformations. The best known chain of solutions
is that of the D$p$-branes, with $p=0,\ldots,8$ in 10 dimensions. They
belong to two different theories: 10-dimensional type~IIA for $p$ even
and 10-dimensional type~IIB for $p$ odd. All of them preserve $1/2$ of
the supersymmetries available, represent objects with $p$ spatial
worldvolume dimensions and $9-p$ transverse dimensions (Dirichlet
branes), carrying charge associated to the RR $(p+1)$-form
$\hat{C}^{(p+1)}$ whose existence was discovered by Polchinski
\cite{kn:P}, and are related by generalized Buscher type~II T~duality
transformations \cite{kn:BHO,kn:MO}.

Sometimes it is possible to find families of solutions that are, by
themselves, representations of the duality group in the sense that
they are invariant, as families, under the full duality group. This is
the case, for instance, of the SWIP solutions of $N=4,d=4$
supergravity constructed in Ref.~\cite{kn:BKO3,kn:L-TO}. In that case
one can argue that all the solitonic objects of a given type (charged,
stationary, black holes) and preserving a certain amount of
supersymmetry are described by particular solutions, with particular
values of the parameters of that general family.  More interesting
cases are $N=8$, $N=4$ with 22 vector multiplets and general $N=2$
theories, all in $d=4$, but fully general solutions in their
duality-invariant form are not available. A great deal is, however,
known of the solitonic spectrum of the 4-dimensional theories due to
our knowledge of their duality groups (the so-called U~duality group
in the $N=8$ case).  All these theories can be obtained from
10-dimensional theories by compactification (toroidal or more general)
and the compactification of the solitonic 10-dimensional objects gives
rise to 4-dimensional solitonic objects of different kinds, depending
on how the 10-dimensional objects are wrapped in the internal
dimensions and one can study if these objects fill 4-dimensional
duality multiplets.  It has been realized that this is not the case if
one considers only the standard 10-dimensional solitons: the
D$p$-branes, KK monopole, gravitational wave (W), fundamental string
(F1) and solitonic 5-brane (S5) \cite{kn:H,kn:H2,kn:BO'L,kn:OPR}. More
10-dimensional solitons are needed to give rise to all the
4-dimensional solitons predicted by duality and some of the properties
they should exhibit, in particular the dependence of the mass in the
radii of the internal dimensions and the coupling constant, have been
deduced.

In this paper we present candidates for some of the missing
10-dimensional solitons and study them. The key to their construction
is the realization that there are 4-dimensional duality symmetries
which are neither present in 10 dimensions nor are a simple
consequence of reparametrization invariance in the internal
coordinates. These are, in general, S~duality (i.e.~electric-magnetic)
transformations which only exist in certain dimensions and that enable
us to use the mechanism {\it reduction-S~dualization-oxidation} to generate
new solutions in higher dimensions.

Let us consider a familiar example: 5 dimensional gravity compactified
in a circle. The 4-dimensional theory has electric-magnetic duality
and one expects an S~duality symmetric spectrum. However, if we only
considered the 5-dimensional plane wave solution we would only find
electrically charged 4-dimensional solitons. To find the magnetically
charged ones we S~dualize these and, oxidizing the solutions to 5
dimensions we find the Kaluza-Klein (KK) monopole
\cite{kn:So,kn:GrPe}.  In principle, this is a solution one would not
expect in 5 dimensions since it has one dimension necessarily
compactified in a circle.

The solutions we present can be generated in a similar fashion,
exploiting S~dualities present in dimensions lower than 10 and 11 and
have similar properties: there are dimensions that cannot be
decompactified. Somehow this is consistent with the fact that they are
generated using dualities that only exist if some of the dimensions
are compact.

One of the problems raised by the need to consider new 10- and
11-dimensional solutions was that fact that the 10- and 11-dimensional
supersymmetry algebras did not contain central charges associated to
those possible new objects. In our opinion the predictive power of the
supersymmetry algebras has been overestimated and we will propose a
way to include in them these new objects.

The rest of this paper is organized as follows: in
Section~\ref{sec-basicsolutions} we present our family of
T~duality-related solutions whose construction via the {\it
  reduction-S~dualization-oxidation} mechanism is explained in
Section~\ref{sec-construction}. In Section~\ref{sec-dualrelated} we
find other duality-related solutions in 10 and 11 dimensions. In
Section~\ref{sec-masses} we calculate the dependence of the masses of
these objects on compactification radii and coupling constants and in
Section~\ref{sec-Killingspinors} we calculate the Killing spinors of
all the solutions we have presented.  Our conclusions are in
Section~\ref{sec-conclusions}. In Appendix~\ref{sec-sl2z} we derive
the $SL(2,\mathbb{R})/SO(2)$ sigma model from toroidal
compactification and explain how $SL(2,\mathbb{R})$ is broken to
$SL(2,\mathbb{Z})$ and in Appendix~\ref{sec-holomorphic} we briefly
review holomorphic $(d-3)$-brane solutions of the
$SL(2,\mathbb{R})/SO(2)$ sigma model to clarify certain points.


\section{The Basic Family of Solutions}
\label{sec-basicsolutions}

The basic family of solutions are solutions of the type~II
supergravity theories in $d=10$ and are a sort of deformation of the
family of D$p$-brane solutions for $0\leq p\leq 7$. As such, they have
$p+1$ worldvolume coordinates $t,\vec{y}_{p}=(y^{1},\ldots,y^{p})$ and
$9-p$ transverse coordinates. We combine two of them into the complex
coordinate $\omega$ and the remaining $7-p$ we denote by
$\vec{x}_{7-p}=(x^{1},\ldots,x^{7-p})$. The solutions can collectively
be written in the string-frame metric in the form\footnote{For
  convenience, we give the form of the potential to which the
  $p$-brane naturally couples $\hat{C}^{(p+1)}$ and the dual one
  $\hat{C}^{(7-p)}$. In the $p=3$ case, these are the two
  non-vanishing sets of components of the 4-form potential with
  self-dual field strength. (Our conventions are those of
  Ref.~\cite{kn:MO} whose type~II T~duality rules, generalizing those
  of Ref.~\cite{kn:BHO}, we use.) Since the solutions we will be
  dealing with are not asymptotically flat, we do not write explicitly
  the asymptotic values of the scalars (for example, $\hat{\phi}_{0}$
  for the dilaton).}

\begin{equation}
\left\{
\label{eq:typeIIsolutions}
\begin{array}{rcl}
d\hat{s}_{s}^2 & = & 
\left(\displaystyle{\frac{H}{{\cal H}\bar{\cal H}}}\right)^{-1/2}
\left(dt^{2}-d\vec{y}_{p}^{\ 2} \right) 
-\left(H{\cal H}\bar{\cal H}\right)^{1/2}d\omega d\bar{\omega}
-\left(\displaystyle{\frac{H}{{\cal H}\bar{\cal H}}}\right)^{1/2}
d\vec{x}_{7-p}^{\ 2}\, ,\\
& & \\
& & \hspace{-1.6cm}\hat{C}^{(p+1)}{}_{ty^{1}\cdots y^{p}} \,=\,  
(-1)^{\left[\frac{(p+1)}{2}\right]} 
\left( \displaystyle{\frac{H}{{\cal H}\bar{\cal H}}}
\right)^{-1}\, ,\\
& & \\
& & 
\hspace{-1.6cm}
\hat{C}^{(7-p)}{}_{x^{1}\ldots x^{7-p}}
   \,= \, -\displaystyle{ \frac{A}{{\cal H}\bar{{\cal H}}} }\, , \\
& & \\
e^{\hat{\phi}} & = & 
 \left( \displaystyle{\frac{H}{{\cal H}\bar{\cal H}}}
\right)^{\frac{3-p}{4}} \, ,
\end{array}
\right.
\end{equation}

\noindent where we  function ${\cal H}={\cal H}(\omega)$ is a complex, 
holomorphic, function of ${\omega}$, i.e.~$\partial_{\bar{\omega}}
{\cal H}=0$ with the behavior ${\cal H}\sim \frac{1}{2\pi
  i}\log{\omega}$ around $\omega=0$, where we assume the object is
placed.  Its real and imaginary parts are

\begin{equation}
{\cal H} = A+iH\, .
\end{equation}

These solutions have the same form as the standard D$p$-brane
solutions if we delete everywhere the combination ${\cal H}\bar{\cal
  H}$, but they are clearly different. In particular we can understand
them as having $7-p$ extra isometric directions that should be
considered compact\footnote{It seems difficult (it is perhaps
  impossible) to extend the dependence of the function ${\cal H}$ to
  those coordinates.  Furthermore, the construction procedure {\it
    reduction-S~dualization-oxidation} and the dependence of the
  masses on the radii of those dimensions that we are going to
  calculate later on suggest that those coordinates should be
  compactified on a torus.}. Our goal will be to understand how they
arise, their M~theoretic origin and their supersymmetry properties and
explore the implications of it all.  We will also find other solutions
related by dualities with them or belonging to the same class. Since
we will find that all these solutions preserve a half of the
symmetries, we are going to argue that they describe the long range
fields of elementary, non-perturbative objects of string theory and we
will calculate their masses.


\section{Construction of the Solutions}
\label{sec-construction}

The solutions (\ref{eq:typeIIsolutions}) can be obtained by successive
T~duality transformations in worldvolume directions of the $p=7$
solution. The $p=7$ solution is nothing but the type~IIB solitonic
7-brane (S7) that was obtained by S~duality from the D7-brane and
called $Q7$-brane in Ref.~\cite{kn:MO}.  The worldvolume directions
are transformed into transverse isometric directions that should be
considered compact\footnote{This is somewhat analogous to what happens
  in the well-known duality between the solitonic fivebrane S5 and the
  KK monopole in which a transverse direction of the S5 is T~dualized
  into an isometric, compact, direction of the KK monopole.}. Thus, we
obtain a chain of T~dual solutions of both type~II theories.

There is an alternative way of constructing these solutions that also
helps to understand them. Let us consider a piece of the
10-dimensional type~II supergravity theories in which we only keep the
metric, the dilaton and the field strength $\hat{G}^{(8-p)}$ of the RR
$(7-p)$-form $\hat{C}^{(7-p)}$. The action is

\begin{equation}
\hat{S} = 
\int d^{10}\hat{x}\sqrt{|\hat{g}|}\,
\left\{e^{-2\hat{\phi}}\left[ \hat{R} -4(\partial\hat{\phi})^{2}\right] 
+{\textstyle\frac{(-1)^{7-p}}{2\cdot (8-p)!}} 
\left(\hat{G}^{(8-p)}\right)^{2}\right\}\, .  
\end{equation}

\noindent Now, let us compactify it over a
$(7-p)$-torus using a simplified Kaluza-Klein Ansatz that only takes
into account the volume modulus of the internal torus, the dilaton
(both rewritten in terms of two convenient scalars $\varphi$ and
$\eta$), the internal volume mode of the RR $(7-p)$-form, $a$ and the
$(3+p)$-dimensional Einstein metric $g_{\mu\nu}$:

\begin{equation}
\left\{
\begin{array}{rcl}
d\hat{s}^{2} & = & e^{\frac{1}{2}\varphi 
+\frac{1}{2}\sqrt{\frac{7-p}{p+1}}\eta}
g_{\mu\nu}dx^{\mu}dx^{\nu} 
-e^{-\frac{1}{2}\varphi 
+\frac{1}{2}\sqrt{\frac{p+1}{7-p}}\eta}
d\vec{x}_{7-p}^{\ 2} \, ,\\
& & \\
\hat{C}^{(7-p)}{}_{x^{1}\ldots x^{7-p}} & = & a\, ,\\
& & \\
e^{\hat{\phi}} & = & 
e^{\frac{p-3}{4}\varphi 
+\frac{7-p}{4}\sqrt{\frac{p+1}{7-p}}\eta}\, .\\
\end{array}
\right.
\end{equation}

\noindent After some straightforward calculations one obtains, in
all cases, the reduced action

\begin{equation}
\label{eq:thereducedaction}
S = 
\int d^{p+3}x \sqrt{|g|}\,
\left\{R +{\textstyle\frac{1}{2}}
\frac{\partial\tau\partial\bar{\tau}}{(\Im {\rm m} \tau)^{2}}
+{\textstyle\frac{1}{2}}(\partial\eta)^{2}
\right\}\, ,  
\end{equation}

\noindent where 

\begin{equation}
\tau = a+ie^{-\varphi}\, ,
\end{equation}

\noindent i.e.~gravity coupled to an $SL(2,\mathbb{R})/SO(2)$ sigma model 
parametrized in the standard form by the complex scalar (sometimes
known as {\it axidilaton} although here this name could be misleading
since in some cases ($p=3$) the string dilaton simply does not
contribute to it) $\tau$ and another scalar, $\eta$, decoupled from
$\tau$.  In the $p=7$ case ($d=10$) this is the well-known piece of
the type~IIB supergravity action. In lower dimensions, it is
integrated in much bigger sigma models associated to much bigger
U-duality groups\footnote{In $d=6$ dimensions, this model was studied
  in Ref.~\cite{kn:BBO} and in $d=8$ it was studied in
  Ref.~\cite{kn:ILPT}.} but it is a most interesting part of it.

There is a very general solution of this model

\begin{equation}
\left\{ 
\begin{array}{rcl}
ds^{2} & = & dt^{2}-d\vec{y}_{p}^{\ 2} -Hd\omega d\bar{\omega}\, ,\\
& & \\
\tau & = & {\cal H}\, ,\\
& & \\
\eta & = & 0\, ,\\
\end{array}
\right.
\end{equation}

\noindent with $\partial_{\bar{\omega}}{\cal H}=0$. In $d=10$ ($p=7$)
this is just the general D7-brane solution. Choosing ${\cal H}\sim
\log{\omega}$ we get the single D7-brane solution. In lower
dimensions, these solutions are just compactifications of the standard
general D$p$-brane solution in which we have assumed that the harmonic
function only depends on two transverse directions ($\omega$) and we
have dualized the RR $(p+1)$-potential, giving rise to the real part
of ${\cal H}$. Thus, this is a well-known solution.

We can now perform an $SL(2,\mathbb{R})$ duality rotation of this
solution\footnote{Continuous duality symmetries are usually broken to
  their discrete subgroups, for instance $SL(2,\mathbb{R})$ is usually
  broken to $SL(2,\mathbb{Z})$. This can be clearly seen in the case
  in which the $SL(2,\mathbb{R})/SO(2)$ sigma model originates in a
  toroidal compactification and is explained in
  Appendix~\ref{sec-sl2z}. In other cases one has to study the
  quantization of charges to arrive to the same conclusion.  We will
  loosely use the continuous of the discrete form of the duality group
  in the understanding that in some contexts only the discrete one is
  really a symmetry of the theory.}  $\tau\rightarrow -1/\tau$, since
this is a symmetry of the dimensionally reduced action\footnote{In
  general, it is only a symmetry of the equations of motion of the
  complete, untruncated, type~II theory.} that leaves the Einstein
metric invariant.  This is not a symmetry of the 10-dimensional action
and one really needs extra compact dimensions to establish it. The
resulting solutions\footnote{In Appendix~\ref{sec-holomorphic} we
  discuss these general solutions and in which sense they are new.  We
  stress that we are considering only the choice holomorphic function
  ${\cal H}\sim \frac{1}{2\pi i}\log{\omega}$.}

\begin{equation}
\label{eq:otherform}
\left\{ 
\begin{array}{rcl}
ds^{2} & = & dt^{2}-d\vec{y}_{p}^{\ 2} -Hd\omega d\bar{\omega}\, ,\\
& & \\
\tau & = & -1/{\cal H}\, ,\\
& & \\
\eta & = & 0\, ,\\
\end{array}
\right.
\end{equation}

\noindent are nothing but the solutions  Eqs.~(\ref{eq:typeIIsolutions}) 
reduced according to the above KK Ansatz

What we are doing here is similar to what one does in standard KK
theory: reducing to 4 dimensions the 5-dimensional pp wave one obtains
the electric, extreme KK black hole. Since the $d=4$ theory has
electric-magnetic duality as a symmetry, one can find the magnetic,
extreme KK black hole and then uplift it to $d=5$ to find the KK
monopole \cite{kn:So,kn:GrPe} that has a special isometric direction
that cannot be decompactified. The symmetry between the pp wave and
the KK monopole cannot be established without assuming one compact
direction. It is only natural, by analogy, to consider here that the
dimensions that we have compactified cannot be decompactified after
the duality transformations. We will support this assumption not by
geometrical arguments but calculating the masses of these objects and
finding its dependence on the radii of those dimensions.


\section{Duality-related Solutions and M-theoretic Origin}
\label{sec-dualrelated}

Since we are dealing with many new solutions, we first propose to
denote them by ``$Dp_{i}$'' where ``$p+1$'' is the worldvolume and
``i'' is the number of isometric directions. According to this
notation, the solutions described by Eq.~(\ref{eq:typeIIsolutions})
are in the $p=7$ case $D7_{0}$ (the type~IIB S~dual of the D7-brane,
called Q7 in Ref.~\cite{kn:MO}), $D6_{1}$ for $p=6$, and $D5_{2}$,
$D4_{3}$, $D3_{4}$, $D2_{5}$, $D1_{6}$, $D0_{7}$ for the remaining
cases.

For all the type~IIB solutions in the class (\ref{eq:typeIIsolutions})
we can find an S~dual using the 10-dimensional type~IIB S~duality
symmetry. While in the $p=7$ case the S~dual solution is just the
well-known D7-brane, and in the $p=3$ case the solution is self-dual,
in the $p=5,1$ cases we find genuinely new solutions. For $D5_{2}$ we
get a solution which is a deformation of the solitonic fivebrane, and
we call $S5_{2}$

\begin{equation}
{\bf S5_{2}}\,\,\,\,
\left\{
\label{eq:p=5Sdual}
\begin{array}{rcl}
d\hat{s}_{s}^2 & = & 
dt^{2}-d\vec{y}_{5}^{\ 2} -Hd\omega d\bar{\omega}
-\displaystyle{\frac{H}{{\cal H}\bar{\cal H}}}
d\vec{x}_{2}^{\ 2}\, ,\\
& & \\
& & \hspace{-1.6cm}\hat{\cal B}_{x^{1}x^{2}} \,=\,  
-\displaystyle{ \frac{A}{{\cal H}\bar{\cal H}} }\, , \\
& & \\
& & \hspace{-1.6cm}\hat{\tilde{\cal B}\, }_{ty^{1}\cdots y^{5}} \,=\,  
\left(\displaystyle{ \frac{H}{{\cal H}\bar{\cal H}} }\right)^{-1}\, , \\
& & \\
e^{\hat{\phi}} & = & 
 \left( \displaystyle{\frac{H}{{\cal H}\bar{\cal H}}}
\right)^{\frac{1}{2}} \, ,
\end{array}
\right.
\end{equation}

\noindent  and for $D1_{6}$, we get a sort of deformation of the fundamental
string solution that we call $F1_{6}$

\begin{equation}
{\bf F1_{6}}\,\,\,\,
\left\{
\label{eq:p=1Sdual}
\begin{array}{rcl}
d\hat{s}_{s}^2 & = & 
\left(\displaystyle{\frac{H}{{\cal H}\bar{\cal H}}}\right)^{-1}
\left(dt^{2}-dy^{2} -Hd\omega d\bar{\omega}\right)
-d\vec{x}_{6}^{\, 2}\, ,\\
& & \\
& & \hspace{-1.6cm}\hat{\cal B}_{ty} \,=\,  
-\left(\displaystyle{ \frac{H}{{\cal H}\bar{\cal H}} }\right)^{-1}\, , \\
& & \\
& & \hspace{-1.6cm}\hat{\tilde{\cal B}\, }_{x^{1}\cdots x^{6}} \,=\,  
\displaystyle{ \frac{A}{{\cal H}\bar{\cal H}} }\, , \\
& & \\
e^{\hat{\phi}} & = & 
 \left( \displaystyle{\frac{H}{{\cal H}\bar{\cal H}}}
\right)^{-\frac{1}{2}} \, .
\end{array}
\right.
\end{equation}

These two solutions only have non-trivial common sector NSNS fields
and therefore they are also solutions of the heterotic string
effective field theory. We can also understand these solutions by
appealing to the existence in both cases of a reduced action of the
form Eq.~(\ref{eq:thereducedaction}) that arises from the
10-dimensional actions

\begin{equation}
\hat{S} = 
\int d^{10}\hat{x}\sqrt{|\hat{g}|}\,
e^{-2\hat{\phi}}\left[ \hat{R} -4(\partial\hat{\phi})^{2}
+{\textstyle\frac{1}{2\cdot 3!}} \hat{H}^{2}\right]\, ,  
\end{equation}

\noindent and 

\begin{equation}
\hat{S} = 
\int d^{10}\hat{x}\sqrt{|\hat{g}|}\,
\left\{e^{-2\hat{\phi}}\left[ \hat{R} -4(\partial\hat{\phi})^{2}\right] 
+{\textstyle\frac{1}{2\cdot 7!}} e^{2\hat{\phi}}
\hat{\tilde{H}}{}^{2}\right\}\, ,  
\end{equation}

\noindent where $\hat{H}$ is the NSNS 3-form field strength and 
$\hat{\tilde{H}}=e^{2\hat{\phi}}\,\,{}^{\star}\hat{H}$ is the dual 7-form
field strength. Reducing the first action to 8 dimensions with the Ansatz

\begin{equation}
\left\{
\begin{array}{rcl}
d\hat{s}^{2} & = &  e^{\frac{1}{\sqrt{3}}\eta}
g_{\mu\nu}dx^{\mu}dx^{\nu} 
-e^{-\varphi}d\vec{x}_2^{\ 2}\, ,\\
& & \\
\hat{B}_{x^{1}x^{2}} & = & a\, ,\\
& & \\
e^{\hat{\phi}} & = & e^{\frac{\sqrt{3}}{2}\eta
-\frac{1}{2}\varphi} \, ,\\
\end{array}
\right.
\end{equation}

\noindent and the second action down to 4 dimensions with the Ansatz 

\begin{equation}
\left\{
\begin{array}{rcl}
d\hat{s}^{2} & = & e^{\varphi}g_{\mu\nu}dx^{\mu}dx^{\nu}
-e^{\frac{1}{\sqrt{3}}\eta}d\vec{x}_{6}^{\ 2} \, ,\\
& & \\
\hat{\tilde{B}\, }_{x^{1}\cdots x^{6}} & = & - a\, ,\\
& & \\
e^{\hat{\phi}} & = & e^{\frac{1}{2}\varphi 
+\frac{\sqrt{3}}{2}\eta}  \, ,\\
\end{array}
\right.
\end{equation}

\noindent we get in both cases Eq.~(\ref{eq:thereducedaction}) in 8 and 4 
dimensions.

As for the M-theoretic origin of the type~IIA solutions, they can be
derived from the following 11-dimensional solutions through
compactification of the 11th dimension ($z$): a $pp$ wave with 7 extra
isometries

\begin{equation}
{\bf WM_{7}}\,\,\,\,
d\hat{\hat{s} }{}^{2}=-2dtdz-\frac{H}{{\cal H}\bar{\cal H}} dz^{2}
-{\cal H}\bar{\cal H}d\omega d\bar{\omega} -d\vec{x}^{\ 2}_{7}\, ,
\end{equation}

\noindent a deformation of the M2-brane

\begin{equation}
{\bf M2_{6}}\,\,\,\,
\left\{
\begin{array}{rcl}
d\hat{\hat{s}}{}^{2} & = &  
\left( \displaystyle{\frac{H}{{\cal H}\bar{\cal H}}}\right)^{-2/3}
(dt^{2}-d\vec{y}^{\ 2}_{2}) 
-H^{1/3}\left({\cal H}\bar{\cal H}\right)^{2/3}d\omega d\bar{\omega}
-\left( \displaystyle{\frac{H}{{\cal H}\bar{\cal H}}}\right)^{1/3}
d\vec{x}^{\ 2}_{6}\, ,\\
& & \\
& & 
\hspace{-1.6cm}
\hat{\hat{C}\, }_{ty^{1}y^{2}} = 
-\left( \displaystyle{\frac{H}{{\cal H}\bar{\cal H}}}\right)^{-1}\, ,\\
& & \\
& & 
\hspace{-1.6cm}
\hat{\hat{\tilde{C}\, }} _{x^{1}\cdots x^{6}} = 
\displaystyle{\frac{A}{{\cal H}\bar{\cal H}}}\, ,\\
\end{array}
\right.
\end{equation}

\noindent a deformation of the M5-brane

\begin{equation}
{\bf M5_{3}}\,\,\,\,
\left\{
\begin{array}{rcl}
d\hat{\hat{s}}{}^{2} & = &  
\left( \displaystyle{\frac{H}{{\cal H}\bar{\cal H}}}\right)^{-1/3}
(dt^{2}-d\vec{y}^{\ 2}_{5}) 
-H^{2/3}\left({\cal H}\bar{\cal H}\right)^{1/3}d\omega d\bar{\omega}
-\left( \displaystyle{\frac{H}{{\cal H}\bar{\cal H}}}\right)^{2/3}
d\vec{x}^{\ 2}_{3}\, ,\\
& & \\
& & 
\hspace{-1.6cm}
\hat{\hat{\tilde{C\, }}}_{ty^{1}\cdots y^{5}}  =  
-\left( \displaystyle{\frac{H}{{\cal H}\bar{\cal H}}}\right)^{-1}\, ,\\
& & \\
& & 
\hspace{-1.6cm}
\hat{\hat{C\, }}_{x^{1}x^{2}x^{3}}  =  
-\displaystyle{\frac{A}{{\cal H}\bar{\cal H}}}\, ,\\
\end{array}
\right.
\end{equation}

\noindent and the KK monopole (with no dependence on the 11th dimension)

\begin{equation}
{\bf KK7M}\,\,\,\,
d\hat{\hat{s}}{}^{2} = dt^{2}-d\vec{y}^{\ 2}_{6}
-H(d\omega d\bar{\omega} +dz^{2}) -H^{-1}
\left(dy^{7} -Adz \right)^{2}\, .
\end{equation}

In these four cases we can also trace the origin of the solution to
the existence of a sector like that in Eq.~(\ref{eq:thereducedaction})
in the reduced action of 11-dimensional supergravity. In the purely
gravitational cases, the action Eq.~(\ref{eq:thereducedaction}) can be
derived from the dimensional reduction of the Einstein term alone as
shown in detail in Appendix~\ref{sec-sl2z}. In the second and third
cases, one needs the 6-form or the 3-form dual potential respectively.

In some cases the dimensional reduction of these 11-dimensional
solutions in isometric directions different from $z$ produce new
10-dimensional solutions. In particular, we get two purely
gravitational solutions

\begin{equation}
{\bf W_{6}}\,\,\,\,
d\hat{s}^{2}=-2dtdz - \frac{H}{{\cal H}\bar{\cal H}} dz^{2}
-{\cal H}\bar{\cal H}d\omega d\bar{\omega} -d\vec{x}^{\ 2}_{6}\, ,
\end{equation}

\noindent and the Kaluza-Klein monopole with no dependence in $z$

\begin{equation}
{\bf KK6}\,\,\,\,
d\hat{s}^{2} = dt^{2}-d\vec{y}^{\ 2}_{5}
-H(d\omega d\bar{\omega} +dz^{2}) -H^{-1}
\left(dy^{7} -Adz \right)^{2}\, .
\end{equation}

\begin{figure}[!ht]
\begin{center}
\leavevmode 
\epsfxsize= 6cm 
\epsfysize= 20cm
\epsffile{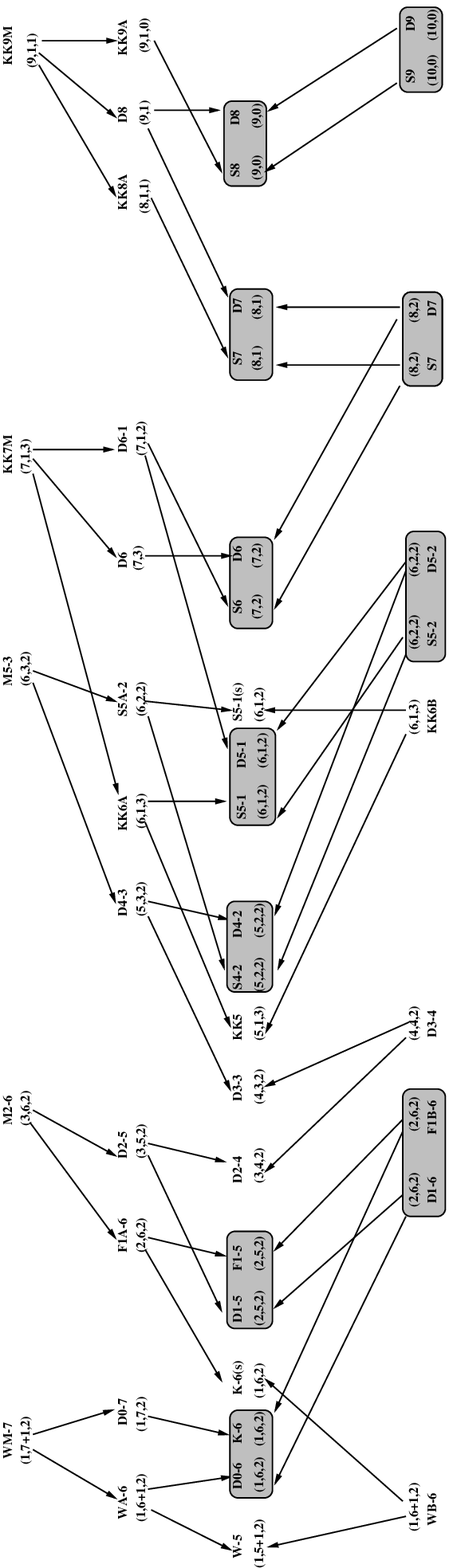}
\caption{\scriptsize Duality relations between 
  KK branes. The numbers in parenthesis represent the worldvolume
  dimension, isometric and transverse directions. The arrows indicate
  dimensional reduction in the corresponding kind of direction. In the
  upper row we represent M-theory KK branes, below 10-dimensional
  type~IIA branes and below them 9-dimensional branes. Type~IIB KK
  branes are in the bottom row. Pairs of branes in boxes are S~duality
  doublets. They are always related to reductions from 11 to 9
  dimensions of the same object in two different orders. Sometimes
  there is an third object with the same numbers as those in a
  doublet, but transforming as a singlet and we denote it with (s).
  \normalsize}
\label{fig:kkbranes}
\end{center}
\end{figure}

In all cases (see Figure~\ref{fig:kkbranes}) we see that whenever we
reduce the same 11-dimensional solution over 2 directions to 9
dimensions and we do it in different order, we get a pair of
9-dimensional solutions that form an $SL(2,\mathbb{R})$
($SL(2,\mathbb{Z})$) doublet and also originate from a type~IIB
$SL(2,\mathbb{R})$ ($SL(2,\mathbb{Z})$) doublet as it must
\cite{kn:BHO}.

\section{Masses}
\label{sec-masses}

The mass of the $Dp_{i}$ solutions can be calculated using S~ and
T~duality rules from the standard $D7$-brane and can be written in a
general formula:

\begin{equation}
M_{Dp_{i}} = \frac{R_{3}\ldots R_{p+2}(R_{p+3}\ldots R_{9})^{2}}
{g^{3}\ell_{s}^{p+2i+1}}\, .  
\end{equation}

The masses of the NSNS solutions found by S~duality from the $D5_{2}$ and
the $D1_{6}$ are

\begin{equation}
  \begin{array}{rcl}
M_{S5_{2}} & =   & 
{\displaystyle
\frac{R_{3}\ldots R_{7} (R_{8}R_{9})^{2}}{g^{2}\ell_{s}^{10}}}\, ,\\
& & \\
M_{F1_{6}} & = &   
{\displaystyle
\frac{R_{3} (R_{4} \ldots R_{9})^{2}}{g^{4}\ell_{s}^{14}}}\, .\\
\end{array}
\end{equation}

The masses of the 11-dimensional objects from which the type~IIA
objects can be derived can be calculated using the relations between
the 11-dimensional Planck length $\ell_{\rm Planck}^{(11)}$ and the
radius of the 11th dimension\footnote{$R_{11}$ is the conventional
  name in the literature. Here we use $R_{m}$ for the radius of the
  coordinate $x^{m}$.} $R_{10}$ and the type~IIA string coupling
constant $g_{A}$ and the string length $\ell_{s}$ $\ell_{\rm
  Planck}{}^{(11)} =2\pi \ell_{s} g_{A}^{1/3}$ and $R_{10}=\ell_{s}
g_{A}$:

\begin{equation}
\begin{array}{rcl}
M_{M2_{6}} & = & {\displaystyle\frac{R_{3}R_{4}(R_{5}\ldots R_{10})^{2}}
{({}^{-}\!\!\!\!\ell_{\rm Planck}^{(11)})^{15}}}\, ,\\
& & \\
M_{M5_{3}} & = & {\displaystyle\frac{R_{3}\ldots 
R_{6}(R_{7}R_{8}R_{9})^{2}R_{10}}
{({}^{-}\!\!\!\!\ell_{\rm Planck}^{(11)})^{12}}}
\, .\\
\end{array}
\end{equation}

\noindent where ${}^{-}\!\!\!\!\ell_{\rm Planck}^{(11)}$ is the reduced
11-dimensional Planck length ${}^{-}\!\!\!\!\ell_{\rm Planck}^{(11)}
=\ell_{\rm Planck}^{(11)}/2\pi$. 

These expressions should be compared with the well-known expression of
the mass of the 11-dimensional KK monopole $KK7M$ when the special
isometric direction is $x^{10}$

\begin{equation}
M_{KK7M} = \frac{R_{4}\ldots R_{9} R_{10}^{2}}
{({}^{-}\!\!\!\!\ell_{\rm Planck}^{(11)})^{9}}\, ,
\end{equation}

\noindent or the 10-dimensional KK monopole $KK6$ ($A$ or $B$) when 
the special isometric direction is $x^{9}$

\begin{equation}
M_{KK6} = \frac{R_{4}\ldots R_{8} R_{9}^{2}}{g^{2}\ell_{s}^{8}}\, .
\end{equation}

\noindent In both cases the mass is not simply proportional to the
volume of the brane which is assumed wrapped on a torus but depends
quadratically on the radius of the special isometric direction.  The
same happens to the masses of all the $Dp_{i}$ branes: they depend
quadratically on the radii of the directions that we have argued are
isometric, which supports our assumption.

Apart from the dependence on the radii we see that in general these
objects are highly non-perturbative since their masses are
proportional to $g^{-3}$ and $g^{-4}$ except for $S5_{2}$, whose mass
goes like $g^{-2}$, as for any standard solitonic object.

The momentum of the $WM_{7}$ solution is 

\begin{equation}
M_{WM_{7}} =\frac{(R_{3}\ldots R_{9})^{2} R_{10}^{3}}
{({}^{-}\!\!\!\!\ell_{\rm Planck}^{(11)})^{18}}\, .  
\end{equation}


\section{Killing Spinors and Unbroken Supersymmetries}
\label{sec-Killingspinors}

It is important to find the amount of supersymmetry preserved by our
solutions since, if they preserve less than one half of the total
supersymmetry available, one could argue that they correspond to
composite objects. Since all these solutions are related by S~and
T~duality transformations to the $D7$-brane, which preserves exactly
$1/2$ of the supersymmetries, it is to be expected that they will do
so as well. Nevertheless, a direct calculation of the Killing spinors
should always be performed since it will confirm our expectations and
it will also provide us with projectors that will help us to associate
the solutions to central charges in the supersymmetry algebra and
therefore to identify them with supersymmetric states in the string
spectrum.

We first calculate the Killing spinors of the $Dp_{i}$ family of
solutions with the obvious choice for the Vielbein
basis\footnote{Underlined indices are world indices and non-underlined
  indices are tangent space indices. They take values in the ranges
  $i=0,1,\ldots,p$ , $m=p+1,\ldots,7$.}

\begin{equation}
e_{\underline{i}}^{\ i}= 
\left(\frac{{\cal H}\bar{{\cal H}}}{H}\right)^{1/4}\, , 
\hspace{1cm}
e_{\underline{m}}^{\ m}=
\left(\frac{{\cal H}\bar{{\cal H}}}{H}\right)^{-1/4}\, ,
\hspace{1cm}
e_{\underline{8}}^{\ 8}=e_{\underline{9}}^{\ 9}=
    \left(\frac{{\cal H}\bar{{\cal H}}}{H}\right)^{1/4}\, H^{1/2}\, .
\end{equation}


For the Type~IIA solutions we use the supersymmetry transformation
rules for the gravitino and dilatino which, in the purely bosonic
background we are considering, take the form\footnote{Our type~IIA
  spinors are full 32-component Majorana spinors.}

\begin{equation}
\left\{
\begin{array}{rcl}
\delta_{\hat{\epsilon}}\hat{\psi}_{\hat{\mu}} 
& = &
\left[ \partial_{\hat{\mu}} 
-\textstyle\frac{1}{4}\not\!\hat{\omega}_{\hat{\mu}} +
\textstyle\frac{i}{8}\textstyle\frac{1}{(8-p)!}
e^{\hat{\phi}}\not\!\hat{G}^{(8-p)} \hat{\Gamma}_{\hat{\mu}}
(-\hat{\Gamma}_{11})^{\frac{8-p}{2}}\right]\hat{\epsilon}\, ,\\
& & \\
\delta_{\hat{\epsilon}}\hat{\tau} 
& = &
\left[\not\!\partial\hat{\phi} + 
\textstyle\frac{i}{4}\textstyle\frac{p-3}{(8-p)!}
e^{\hat{\phi}}\not\!\hat{G}^{(8-p)}
(-\hat{\Gamma}_{11})^{\frac{8-p}{2}}\right]\hat{\epsilon}\, ,\\
\end{array}
\right.
\end{equation}

\noindent Imposing the 
vanishing of dilatino transformation rule we obtain the following
constraint in the Killing spinor:

\begin{equation}
  \label{constraintIIA}
  \left[1-i\hat{\Gamma}^{p+1}\cdots\hat{\Gamma}^{8}\hat{\Gamma}^{9}
  (-\hat{\Gamma}_{11})^{\frac{8-p}{2}}\right]\hat{\epsilon}=0\, , 
\end{equation}

\noindent or, equivalently

\begin{equation}
  \label{constraintIIA2}
  \left[1-(-1)^{[p/2]} i\hat{\Gamma}^{0}\cdots\hat{\Gamma}^{p}
  (-\hat{\Gamma}_{11})^{\frac{10-p}{2}}\right]\hat{\epsilon}=0\, .
\end{equation}

\noindent This constraint automatically sets to zero the worldvolume
($t,y^{i}$) and transverse, isometric ($x^{m}$) components of the
supersymmetry variation of the gravitino. The remaining transverse
components ($x^{8},x^{9}$) give in {\em all} cases, the following
coupled partial differential equations

\begin{equation}
\left\{
\begin{array}{rcl}
\delta_{\hat{\epsilon}}\hat{\psi}_{\underline{8}}
& = &
\left[\partial_{\underline{8}}
-\textstyle\frac{1}{4}\hat{\Gamma}^{8}\hat{\Gamma}^{9}\, 
\partial_{\underline{9}}\log({\cal H}\bar{\cal H})+
\textstyle\frac{1}{8}\, 
\partial_{\underline{8}}
\displaystyle{\log
\left(\frac{H}{{\cal H}\bar{{\cal H}}}\right)}\right]\hat{\epsilon}=0\, , \\
& & \\
\delta_{\hat{\epsilon}}\hat{\psi}_{\underline{9}}
& = &
\left[\partial_{\underline{9}}
-\textstyle\frac{1}{4}\hat{\Gamma}^{9}\hat{\Gamma}^{8}\, 
\partial_{\underline{8}}\log( {\cal H}\bar{{\cal H}}) +
\textstyle\frac{1}{8}\, 
\partial_{\underline{9}}
\displaystyle{\log
\left(\frac{H}{{\cal H}\bar{{\cal H}}}\right)}\right]\hat{\epsilon}=0\, .\\
\end{array}
\right.
\end{equation}

Now, using the Cauchy-Riemann equations for the holomorphic function
${\cal H}$, i.e.:

\begin{equation}
\partial_{\underline{8}}A =
  +\partial_{\underline{9}}H\, ,
\hspace{1cm}
\partial_{\underline{9}}A =
  -\partial_{\underline{8}}H\, ,
\end{equation}

\noindent we can express $\partial_{\underline{8}}\log({\cal H}\bar{\cal H})$ 
and $\partial_{\underline{9}}\log({\cal H}\bar{\cal H})$ in the following way:

\begin{equation}
\partial_{\underline{9}}\log( {\cal H}\bar{{\cal H}}) =
-2\partial_{\underline{8}}(\arg{\cal H})\, ,
\hspace{1cm}
\partial_{\underline{8}}\log( {\cal H}\bar{{\cal H}}) =
+2\partial_{\underline{9}}(\arg{\cal H})\, ,
\end{equation}

\noindent and the Killing spinor equations are easily seen to be solved by

\begin{equation}
\left\{
\begin{array}{c}
\left[1-(-1)^{[p/2]} i\hat{\Gamma}^{0}\cdots\hat{\Gamma}^{p}
(-\hat{\Gamma}_{11})^{\frac{10-p}{2}}\right]\hat{\epsilon}_{0}=0\, ,\\
\\
\hat{\epsilon} =
e^{-\frac{1}{2}\arg({\cal H})\, \hat{\Gamma}^{8}\hat{\Gamma}^{9}}
\displaystyle{
\left(\frac{{\cal H}\bar{{\cal H}}}{H}\right)^{1/8}}\hat{\epsilon}_{0}\, .
\end{array}
\right.
\end{equation}

\noindent $\hat{\epsilon}_{0}$ being any constant spinor 
satisfying the above constraint.

  

In the type~IIB cases we use the relevant supersymmetry transformation
laws\footnote{Our type~IIB spinors are pairs (whose indices 1,2 are
  not explicitly shown of 32-component, positive chirality,
  Majorana-Weyl spinors. Pauli matrices act on the indices not shown.}

\begin{equation}
\left\{
\begin{array}{rcl}
\delta_{\hat{\varepsilon}}\hat{\zeta}_{\hat{\mu}} 
& = &
\left[ \partial_{\hat{\mu}} 
-\textstyle\frac{1}{4}\not\!\hat{\omega}_{\hat{\mu}} 
+\textstyle\frac{1}{8}\textstyle\frac{1}{(8-p)!}
e^{\hat{\varphi}}\not\!\hat{G}^{(8-p)}
\hat{\Gamma}_{\hat{\mu}}{\cal P}_{\frac{9-p}{2}}\right]
\hat{\varepsilon}\, ,\\
& & \\
\delta_{\hat{\varepsilon}}\hat{\chi}
& = & 
\left[\not\!\partial\hat{\varphi} 
+\textstyle\frac{1}{4}\textstyle\frac{3-p}{(8-p)!}
e^{\hat{\varphi}}\not\!\hat{G}^{(8-p)}{\cal P}_{\frac{9-p}{2}}\right]
\hat{\varepsilon}\, ,\\
\end{array}
\right.
\end{equation}

\noindent where  ${\cal P}_{n}$ is
 
\begin{displaymath}
  {\cal P}_n\left\{ \begin{array}{ll}
  \sigma^{1}\, , & n \mbox{ even}\, , \\ \\
  i\sigma^{2}\, , & n \mbox{ odd}\, . 
  \end{array}\right.
\end{displaymath}

Proceeding as in the type~IIA case, we find the Killing spinors

\begin{equation}
\left\{
\begin{array}{c}
\left[1+(-1)^{[p/2]}\hat{\Gamma}^{0}\cdots\hat{\Gamma}^{p}
{\cal P}_{\frac{p+1}{2}}\right]\hat{\varepsilon}_{0}=0\, ,\\ 
\\
\hat{\varepsilon} = e^{-\frac{1}{2}\arg({\cal H})\, 
\hat{\Gamma}^{8}\hat{\Gamma}^{9}}
\displaystyle{
\left(\frac{{\cal H}\bar{{\cal H}}}{H}\right)^{1/8}}
\hat{\varepsilon}_{0}\, ,\\
\end{array}
\right.
\end{equation}

\noindent where, now, $\hat{\varepsilon}_{0}$ is any pair of constant
positive-chirality Majorana-Weyl spinors satisfying the above constraint.

The Killing spinors of the $S5_{2}$ and the $F1_{6}$ can be found in a similar fashion and are, respectively 

\begin{displaymath}
\left\{
\begin{array}{c}
\left[1-\hat{\Gamma}^{6}\hat{\Gamma}^{7}\hat{\Gamma}^{8}\hat{\Gamma}^{9}
\sigma^3\right]\hat{\varepsilon}_{0}=0\, ,\\ 
\\
\hat{\varepsilon} = e^{-\frac{1}{2}\arg({\cal H})\, 
\hat{\Gamma}^{8}\hat{\Gamma}^{9}}
\hat{\varepsilon}_{0}\, ,\\
\end{array}
\right.
\end{displaymath}

\noindent and

\begin{displaymath}
\left\{
\begin{array}{c}
\left[1+\hat{\Gamma}^{0}\hat{\Gamma}^{1}\sigma^3
\right]\hat{\varepsilon}_{0}=0\, ,\\ 
\\
\hat{\varepsilon} = e^{-\frac{1}{2}\arg({\cal H})\, 
\hat{\Gamma}^{8}\hat{\Gamma}^{9}}
\displaystyle{
\left(\frac{{\cal H}\bar{{\cal H}}}{H}\right)^{1/4}}
\hat{\varepsilon}_{0}\, ,\\
\end{array}
\right.
\end{displaymath}

Before discussing these results it is worth finding the Killing
spinors of the 11-dimensional solutions. The only relevant
supersymmetry transformation rule is that of the gravitino, which with
our conventions is:

\begin{equation}
\delta_{\hat{\hat{\epsilon}}}\hat{\hat{\psi}}_{\hat{\hat{\mu}}}
=\left[
2\partial_{\hat{\hat{\mu}}}
-\textstyle\frac{1}{2}\not\!\hat{\hat{\omega}}_{\hat{\hat{\mu}}}+
\textstyle\frac{i}{144}\left(
\hat{\hat{\Gamma}}{}^{\hat{\hat{\alpha}}\hat{\hat{\beta}}\hat{\hat{\gamma}}
\hat{\hat{\delta}}}_{\ \ \ \ \ \hat{\hat{\mu}}}
-8\hat{\hat{\Gamma}}{}^{\hat{\hat{\beta}}\hat{\hat{\gamma}}\hat{\hat{\delta}}}
\hat{\hat{\eta}}_{\hat{\hat{\mu}}}{}^{\ \hat{\hat{\alpha}}}\right)
\hat{\hat{G}}_{\hat{\hat{\alpha}}\hat{\hat{\beta}}\hat{\hat{\gamma}}
\hat{\hat{\delta}}}
\right]\hat{\hat{\epsilon}}\, .
\end{equation}

In the obvious Vielbein basis we find, for the $WM_{7}$ solution

\begin{equation}
\left\{
\begin{array}{l}
\left[1-\hat{\hat{\Gamma}}{}^{0}\hat{\hat{\Gamma}}{}^{10}\right]
\hat{\hat{\epsilon}}_{0}=0\, , \\ 
\\
\hat{\hat{\epsilon}}=e^{-\frac{1}{2} \arg({\cal H})
\hat{\hat{\Gamma}}{}^{8}\hat{\hat{\Gamma}}{}^{9}}
\left(\displaystyle{\frac{{\cal H}\bar{\cal H}}{H}}\right)^{1/4}
\hat{\hat{\epsilon}}_{0}\, .\\
\end{array}
\right.
\end{equation}

\noindent for the $M2_{6}$ solution

\begin{equation}
\left\{
\begin{array}{l}
\left[1+i\hat{\hat{\Gamma}}{}^{0}\hat{\hat{\Gamma}}{}^{1}
\hat{\hat{\Gamma}}{}^{2}
\right]\hat{\hat{\epsilon}}_{0}=0\, , \\ 
\\
\hat{\hat{\epsilon}} =e^{-\frac{1}{2} \arg({\cal H})
\hat{\hat{\Gamma}}{}^{8}\hat{\hat{\Gamma}}{}^{9}}
\left(\displaystyle{\frac{{\cal H}\bar{\cal H}}{H}}\right)^{1/6}
\hat{\hat{\epsilon}}_{0}\, ,\\
\end{array}
\right.
\end{equation}

\noindent for the $M5_{3}$ solution

\begin{equation}
\left\{
\begin{array}{l}
\left[1-\hat{\hat{\Gamma}}{}^{0}\cdots\hat{\hat{\Gamma}}{}^{4}
\hat{\hat{\Gamma}}{}^{10}\right]\hat{\hat{\epsilon}}_{0}=0\, , \\ 
\\
\hat{\hat{\epsilon}}=e^{-\frac{1}{2} \arg({\cal H})
\hat{\hat{\Gamma}}{}^{8}\hat{\hat{\Gamma}}{}^{9}}
\left(\displaystyle{\frac{{\cal H}\bar{\cal H}}{H}}\right)^{1/12}
\hat{\hat{\epsilon}}_{0}\, ,\\
\end{array}
\right.
\end{equation}

\noindent and for the $KK7M$ solution, as it is well known,
the Killing spinor is any constant spinor $\hat{\hat{\epsilon}}_{0}$
satisfying the constraint

\begin{equation}
\left[1+i\hat{\hat{\Gamma}}{}^{0}\cdots\hat{\hat{\Gamma}}{}^{6}
\right]\hat{\hat{\epsilon}}_{0}=0\, .
\end{equation}

In all cases one can see that these solutions preserve one half of the 
supersymmetries.




\section{Conclusions}
\label{sec-conclusions}

In this paper we have presented new 10-dimensional solutions of the
type~IIB theories that can be thought of as having a certain number of
isometric, compact, dimensions, that cannot be decompactified (one
could say that these are really solutions of lower-dimensional
theories) and which we have referred generically to as ``KK-branes''.
We have described how they can be obtained via the {\it
  reduction-S~dualization-oxidation} which could explain why some of
the directions have to be compactified in circles since S~duality only
exists in the compactified theory. Furthermore, we have computed the
masses of these solutions and we have found that they depend on the
square of the radii of the directions that we have identified as
compact, just as it happens in the KK monopole case, which is
consistent with our identification. The mass formula are also
coincident with what is needed to complete the U~duality invariant
spectrum of $N=8,d=4$ supergravity \cite{kn:H2,kn:BO'L,kn:OPR}.  It
has also been recently argued that the presence of certain KK-branes
is necessary to explain from the M~theory point of view the existence
of some massive/gauged type~II supergravities in lower dimensions
\cite{kn:A-AMO}.

Perhaps the only element that does not seem to fit in the picture we
are putting forward is the supersymmetry algebra since there seems to
be no place in it for the new objects. For the sake of concreteness
we will focus in the 11-dimensional supersymmetry algebra
(``M~algebra'') but the problems and the solutions we propose can be
applied in the obvious way to other cases.

The M~algebra is usually written, up to convention-dependent numerical
factors $c, c_{n}$, in the form\footnote{See e.g.~\cite{kn:Tow6}.}

\begin{equation}
\left\{Q^{\alpha},Q^{\beta}\right\} 
= c \left(\Gamma^{a}{\cal C}^{-1}\right)^{\alpha\beta}P_{a}
+{\textstyle\frac{c_{2}}{2}}
\left(\Gamma^{a_{1}a_{2}}{\cal C}^{-1}\right)^{\alpha\beta}
{\cal Z}^{(2)}_{a_{1}a_{2}}
+{\textstyle\frac{c_{5}}{5!}}
\left(\Gamma^{a_{1}\cdots a_{5}}{\cal C}^{-1}\right)^{\alpha\beta}
{\cal Z}^{(5)}_{a_{1}\cdots a_{5}}\, .
\end{equation}

A lightlike component of the momentum is then associated to the
gravitational waves moving in that direction, the spatial components
of ${\cal Z}^{(2)}$ and ${\cal Z}^{(5)}$ are associated respectively
to $M2$- and $M5$-branes wraped in those directions. The timelike
components have more complicated interpretations: in the ${\cal
  Z}^{(5)}$ case, they are associated to the KK monopole in a
complicated way and in the ${\cal Z}^{(2)}$ case they are associated
to an object that we would call the $KK9$-brane of which we only know
that it should give the $D8$-brane upon dimensional reduction.  All
these objects break (preserve) a half of the available supersymmetries
and strict relations between their masses and charges can be derived
from the algebra.

Clearly the M~algebra contains a good deal of information about the
solitons of the theory that realizes it (11-dimensional supergravity
or M~theory). However, it is clear that it does not contain all the
information about them. To start with, it does not tell us why some
branes are fundamental and some are solitonic, it does not tell us why
some objects exist in the uncompactified theory (the wave, $M2$ and
$M5$) while other objects only exist when one dimension is
compactified in a circle (the KK monopole and the $KK9$-brane).
Furthermore, all solitonic objects should be associated to spacelike
components of central charges: that is the result we would always get
if we performed the calculation. All this is not so surprising: the
M~algebra is not derived from the theory and their solutions but just
by imposing consistency of the possible central charges. If we were
able to derive the algebra from M~theory and its solitonic solutions,
the central charges would be associated to specific objects and we
would know whether they have compact dimensions or not. Since we do
know many things about the solitonic solutions, we can try to reflect
what we know in a form of the M~algebra mathematically consistent and
then we can check if the results are consistent with dualities.

To start with, we consider the M~algebra with the most general central
extensions allowed:

\begin{equation}
\left\{Q^{\alpha},Q^{\beta}\right\} 
= c \left(\Gamma^{a}{\cal C}^{-1}\right)^{\alpha\beta}P_{a}
+\sum_{n=2,5,6,9,10}{\textstyle\frac{c_{n}}{n!}}
\left(\Gamma^{a_{1}\cdots a_{n}}{\cal C}^{-1}\right)^{\alpha\beta}
{\cal Z}^{(n)}_{a_{1}\cdots a_{n}}\, .
\end{equation}

We know the wave is associated to $P$, the $M2$-brane to ${\cal
  Z}^{(2)}$ and the M5-brane to ${\cal Z}^{(5)}$. We also know
\cite{kn:BJO2} that the KK monopole is a sort of 6-brane with one of
the 4 possible transverse dimensions wrapped in a circle. We are going
to reflect this fact by writing, instead of just the ${\cal Z}^{(6)}$
term as above, the term

\begin{equation}
{\textstyle\frac{c_{6}}{6!}}
\left(\Gamma^{a_{1}\cdots a_{6}}{\cal C}^{-1}\right)^{\alpha\beta}
{\cal Z}^{(7)}_{a_{1}\cdots a_{6} a_{7}} k^{a_{7}}\, ,  
\end{equation}

\noindent where $k^{a}$ is a vector pointing in the compact direction. 

We also know that the $KK9$-brane (or $M9$-brane) \cite{kn:BvdS} has 9
spacelike worldvolume dimensions one of which is always wrapped on a
circle. We reflect this fact by writing, instead of just the ${\cal
  Z}^{(9)}$ term as above, the term

\begin{equation}
{\textstyle\frac{c_{9}}{9!}}
\left(\Gamma^{a_{1}\cdots a_{9}}{\cal C}^{-1}\right)^{\alpha\beta}
{\cal Z}^{(8)}_{a_{1}\cdots a_{8}} l_{a_{9}}\, ,  
\end{equation}

\noindent where $l_{a}$ is a vector pointing in the direction 
around which the $KK9$-brane is wrapped.

We do not know of any brane associated to ${\cal Z}^{(10)}$ and so we
will not consider it in the M~algebra, which takes the form

\begin{equation}
  \begin{array}{rcl}
\left\{Q^{\alpha},Q^{\beta}\right\} 
& = & c \left(\Gamma^{a}{\cal C}^{-1}\right)^{\alpha\beta}P_{a}
+{\textstyle\frac{c_{2}}{2}}
\left(\Gamma^{a_{1}a_{2}}{\cal C}^{-1}\right)^{\alpha\beta}
{\cal Z}^{(2)}_{a_{1}a_{2}}
+{\textstyle\frac{c_{5}}{5!}}
\left(\Gamma^{a_{1}\cdots a_{5}}{\cal C}^{-1}\right)^{\alpha\beta}
{\cal Z}^{(5)}_{a_{1}\cdots a_{5}}\\
& & \\
& & 
+{\textstyle\frac{c_{6}}{6!}}
\left(\Gamma^{a_{1}\cdots a_{6}}{\cal C}^{-1}\right)^{\alpha\beta}
{\cal Z}^{(7)}_{a_{1}\cdots a_{6} a_{7}} k^{a_{7}}
+{\textstyle\frac{c_{9}}{9!}}
\left(\Gamma^{a_{1}\cdots a_{9}}{\cal C}^{-1}\right)^{\alpha\beta}
{\cal Z}^{(8)}_{a_{1}\cdots a_{8}} l_{a_{9}}\, .\\
\end{array}
\end{equation}

We could certainly write more general central charges by allowing more
vectors to be present in the algebra, meaning allowing objects with
more isometric directions such as the $M2_{6}$ or the $M5_{3}$ branes
presented in this paper. However, considering objects with just one special
isometry will be enough to present our ideas.

Let us now reduce this algebra in one dimension. From each of the
standard central charges we get two central charges in one dimension
less, namely $P,{\cal Z}^{(0)}$ from $P$, ${\cal Z}^{(1)},{\cal
  Z}^{(2)}$ from ${\cal Z}^{(2)}$ and ${\cal Z}^{(4)},{\cal Z}^{(5)}$
from ${\cal Z}^{(5)}$, corresponding to the known reductions of
M~theory solitons: wave and $D0$-brane from the wave, $F1$ and
$D2$-brane from the $M2$-brane and $D4$- and $S5$-brane from the
$M5$-brane. From each of the new charges we have introduced we get
instead three lower dimensional central charges: from the contraction
${\cal Z}^{(7)}k$ associated to the KK monopole we get a ${\cal
  Z}^{(6)}$ associated to the $D6$-brane when $k$ points in the
direction we are reducing, we get a contraction ${\cal Z}^{(6)}k$
associated to the type~IIA KK monopole ($KK6A$) if we reduce on the KK
monopole worldvolume and we get a ${\cal Z}^{(7)}k$ associated to the
$D6_{1}$ (called $KK7A$ in Ref.~\cite{kn:MO}, also studied in
Ref.~\cite{kn:EL} ) if we reduce in a transverse direction. From the
product ${\cal Z}^{(8)}l$ we get a ${\cal Z}^{(8)}$, associated to the
$D8$-brane when we reduce the KK9-brane in the isometric direction $l$
points to, we get a product ${\cal Z}^{(7)}l$ associated to an object
with the same features of the M~theory $KK9$-brane but in one
dimension less and a product ${\cal Z}^{(8)}l$ associated to a
type~IIA spacetime filling $KK9$-brane referred to as $NS-9A$-brane in
Ref.~\cite{kn:BEHvdSHL}. The result is the following form of the
type~IIA supersymmetry algebra:

\begin{equation}
  \begin{array}{rcl}
\left\{Q^{\alpha},Q^{\beta}\right\} 
& = & c \left(\Gamma^{a}{\cal C}^{-1}\right)^{\alpha\beta}P_{a}
+\sum_{n=0,1,4,8}{\textstyle\frac{c_{n}}{n!}}
\left(\Gamma^{a_{1}\cdots a_{n}} \Gamma_{11} 
{\cal C}^{-1}\right)^{\alpha\beta}
{\cal Z}^{(n)}_{a_{1}\cdots a_{n}}\\
& & \\
& & 
+\sum_{n=2,5,6}{\textstyle\frac{c_{n}}{n!}}
\left(\Gamma^{a_{1}\cdots a_{n}} {\cal C}^{-1}\right)^{\alpha\beta}
{\cal Z}^{(n)}_{a_{1}\cdots a_{n}}\\
& & \\
& & 
+{\textstyle\frac{c_{5}}{5!}}
\left(\Gamma^{a_{1}\cdots a_{5}}\Gamma_{11}{\cal C}^{-1}\right)^{\alpha\beta}
{\cal Z}^{(6)}_{a_{1}\cdots a_{5} a_{6}} k^{a_{6}}
+{\textstyle\frac{c_{6}}{6!}}
\left(\Gamma^{a_{1}\cdots a_{6}}{\cal C}^{-1}\right)^{\alpha\beta}
{\cal Z}^{(7)}_{a_{1}\cdots a_{6} a_{7}} l^{a_{7}}\\
& & \\
& & 
+{\textstyle\frac{c_{8}}{8!}}
\left(\Gamma^{a_{1}\cdots a_{8}}{\cal C}^{-1}\right)^{\alpha\beta}
{\cal Z}^{(7)}_{a_{1}\cdots a_{7}} m_{a_{8}}
+{\textstyle\frac{c_{9}}{9!}}
\left(\Gamma^{a_{1}\cdots a_{9}}{\cal C}^{-1}\right)^{\alpha\beta}
{\cal Z}^{(8)}_{a_{1}\cdots a_{8}} n_{a_{9}}\, .\\
\end{array}
\end{equation}

Every known solitonic solution of the type~IIA supergravity theory has
an associated charge in this algebra. If we now reduce again to nine
dimensions we will get the algebra of the massive 9-dimensional
theories presented in Ref.~\cite{kn:MO} with $SL(2,\mathbb{Z})$
covariance.  This is possible only because we have allowed for charges
corresponding to KK-branes in 11 dimensions. To get the same algebra
from the type~IIB side a charge has to be introduced for the $S7$
brane which, even though it does not carry any $SO(2)$ R-symmetry
indices, is not invariant but is interchanged with the $D7$-brane
charge under S~duality. We will present these results elsewhere.


\section*{Acknowledgments}

T.O.~would like to thank Eric Bergshoeff, Joaquim Gomis, Yolanda
Lozano and Patrick Meessen for most useful conversations, the
C.E.R.N.~Theory Division for its hospitality in the first stages and
the Groningen Institute for Theoretical Physics for its warm
hospitality in the last stages of this work and M.~Fern\'andez for her
support. 

The work of E.L.-T.~is supported by a U.A.M.~grant for postgraduate
studies. The work of T.O.~is supported by the European Union TMR
program FMRX-CT96-0012 {\sl Integrability, Non-perturbative Effects,
  and Symmetry in Quantum Field Theory} and by the Spanish grant
AEN96-1655.

\newpage
\appendix

\section{Holomorphic $(d-3)$-Branes}
\label{sec-holomorphic}

In this Appendix we briefly discuss holomorphic $(d-3)$-brane
solutions of the $d$-dimensional $SL(2,\mathbb{R})/SO(2)$ sigma model

\begin{equation}
\label{eq:SL2action}
S = \int d^{d}x \sqrt{|g|}\,
\left\{R +{\textstyle\frac{1}{2}}
\frac{\partial\tau\partial\bar{\tau}}{(\Im {\rm m} \tau)^{2}}
\right\}\, ,  
\end{equation}

\noindent where $\tau$ lives in the complex upper half plane and 
is defined up to modular $PSL(2,\mathbb{Z})$ transformations, so 
multivalued solutions are allowed if the value of $\tau$ changes by 
a modular transformation.

$(d-3)$-brane-type solutions of this model were first considered in
Ref.~\cite{kn:GSVY} in $d=4$. In these dimensions $(d-3)$-branes are
strings. In that reference, the following general solution of the
above model was found\footnote{Here we write the obvious
  generalization to any dimension $d$ (see also
  Refs.~\cite{kn:GGP,kn:MO}).}

\begin{equation}
\label{eq:generalsolution1}
\left\{ 
\begin{array}{rcl}
ds^{2} & = & dt^{2}-d\vec{y}_{(d-3)}^{\ 2} -Hd\omega d\bar{\omega}\, ,\\
& & \\
\tau & = & {\cal H}\, ,\\
\end{array}
\right.
\end{equation}

\noindent where ${\cal H}$ is, in principle, any complex
holomorphic or antiholomorphic function of the complex variable
$\omega$ (i.e.~either $\partial_{\bar{\omega}} {\cal H}=0$ or
$\partial_{\omega} {\cal H}=0$) and $H=\Im{\rm m}({\cal H})$. $H$ is,
therefore, a real harmonic function of the 2-dimensional Euclidean
spacetime transverse to the $(d-2)$-dimensional worldvolume
directions. Only functions with $H\geq 0$ are admissible.

A few remarks are in order here: although $g_{\omega\bar{\omega}}=H$
is in this solution equal to the imaginary part of $\tau$, it does not
transform under $PSL(2,\mathbb{Z})$. Modular invariance of the metric
is, therefore, not an issue. We could have wrongly concluded that in
this solution, the metric is not modular invariant because
$g_{\omega\bar{\omega}} =\Im{\rm m}(\tau)$ but, by definition, it is,
since the metric does not transform under $PSL(2,\mathbb{Z})$.  Then,
the l.h.s.~if that equation does not transform, and the r.h.s.~does,
and we get a new solution (denoted by primes) with

\begin{equation}
\begin{array}{rcl}
\tau^{\prime}(\omega) & = & 
{\displaystyle\frac{a\tau(\omega)+b}{c\tau(\omega)+d}}
={\displaystyle\frac{a{\cal H}+b}{c{\cal H}+d}}
\equiv {\cal H}^{\prime}\, ,\\
& & \\
g_{\omega\bar{\omega}}^{\prime} & = & g_{\omega\bar{\omega}}
=\Im{\rm m}(\tau)
={\displaystyle\frac{\Im{\rm m} (\tau^{\prime})}
{|-c\tau^{\prime}+a|^{2}}}
={\displaystyle\frac{\Im{\rm m} ({\cal H}^{\prime})}
{|-c{\cal H}^{\prime}+a|^{2}}}\, .\\
\end{array}
\end{equation}

We could remove if we wished the extra factor by a conformal
reparametrization: 

\begin{equation}
d\omega^{\prime}= \frac{d\omega}{-c{\cal H}^{\prime}(\omega)+a}\, ,
\end{equation}

\noindent and we then could write again the new solution in a form 
similar to that of the original one Eq.~(\ref{eq:generalsolution1})
but with a new holomorphic function ${\cal
  H}^{\prime}[\omega(\omega^{\prime})]$.  Thus, as in
Ref.~\cite{kn:GSVY} we could have written from the beginning the
general solution in the form

\begin{equation}
\label{eq:generalsolution2}
\left\{ 
\begin{array}{rcl}
ds^{2} & = & dt^{2}-d\vec{y}_{(d-3)}^{\ 2} 
-H|f(\omega)|^{2}d\omega d\bar{\omega}\, ,\\
& & \\
\tau & = & {\cal H}\, ,\\
\end{array}
\right.
\end{equation}

\noindent where $f(\omega)$ is any holomorphic function of $\omega$,
but this function can always be reabsorbed into a holomorphic
coordinate change $\omega^{\prime}=F(\omega)\, ,\,\,\,\, dF/d\omega=
f$ and $\tau(\omega^{\prime})=\tau[F^{-1}(\omega^{\prime})]$.

All this said, it must be acknowledged that, even though modular
invariance of the metric is not an issue, its single-valuedness is.
Since ${\cal H}$ will in general be a multivalued function with
monodromies in $G$, its imaginary part will also be multivalued and it
might be necessary to multiply it by $|f(\omega)|^{2}$, with
$f(\omega)$ multivalued to make $g_{\omega\bar{\omega}}$ single
valued. 

A second remark we can make here is that there exists another form of
the general solution which is manifestly $SL(2,\mathbb{R})$ invariant
without having to invoke coordinate changes to show it:

\begin{equation}
\label{eq:thegeneralsolution}
\left\{ 
\begin{array}{rcl}
ds^{2} & = & dt^{2}-d\vec{y}_{p}^{\ 2} -e^{-2U}d\omega d\bar{\omega}\, ,\\
& & \\
\tau & = & {\cal H}_{1}/{\cal H}_{2}\, ,\\
& & \\
e^{-2U} & = & \Im {\rm m}\left( {\cal H}_{1}\bar{\cal H}_{2}\right)\, ,\\
\end{array}
\right.
\end{equation}

\noindent where ${\cal H}_{1,2}$ are two arbitrary complex 
functions of the complex variable $\omega$ transforming as a doublet
under $SL(2,\mathbb{R})$, i.e.

\begin{equation}
\left(
  \begin{array}{c}
{\cal H}^{\prime}_{1}\\
\\
{\cal H}^{\prime}_{2}\\
  \end{array}
\right)  
=
\left(
  \begin{array}{cc}
a & \,\,\,b\\
& \\
c & \,\,\,d \\
  \end{array}
\right)  
\left(
  \begin{array}{c}
{\cal H}_{1}\\
\\
{\cal H}_{2}\\
  \end{array}
\right)\, ,  
\end{equation}

\noindent both in $\tau$ and in the metric (but $e^{-2U}$ is invariant, 
as it must). The structure of this family is similar to that of the
duality-invariant families of black-hole solutions of pure $N=4,d=4$
supergravity presented in Refs.~\cite{kn:KO,kn:BKO3,kn:L-TO}, closely
related to special geometry objects as discovered in \cite{kn:FKS}.
We can relate this general solution either to the solution
Eq.~(\ref{eq:generalsolution1}) as the particular case ${\cal
  H}_{1}={\cal H}\, ,\,\,\,{\cal H}_{2}=1$ or to  the solution
Eq.~(\ref{eq:generalsolution2}) as the particular case ${\cal
  H}_{1}/{\cal H}_{2}={\cal H}\, ,\,\,\, f={\cal H}_{2}$ since
$\Im{\rm m}({\cal H}_{1}\bar{\cal H}_{2})=|{\cal H}_{2}|^{2} \Im{\rm
  m}({\cal H}_{1}/{\cal H}_{2})$.

All this means that we cannot generate new solutions not in this
classes via $SL(2,\mathbb{R})$ transformations. 

Since all these solutions are equivalent, up to coordinate
transformations, we take now Eq.~(\ref{eq:generalsolution1}) and now
consider the choice of function ${\cal H}$.  First, we have to choose
between holomorphic and anti-holomorphic ${\cal H}$. This choice is
related to the choice between $(d-3)$-branes and anti-$(d-3)$-branes
with opposite charge with respect to the $(d-2)$-form potential dual
to $a$. The impossibility of having ${\cal H}$ depending on both
$\omega$ and $\bar{\omega}$ is due to the impossibility of having
objects with opposite charges in equilibrium.  We opt for holomorphy.

Which holomorphic function should one choose? As usual, the choice has
to be based on local and global conditions. Local conditions are
essentially related to the existence of extended sources (with $(d-3)$
spatial dimensions) at given points in transverse ($\omega$) space
manifold.  Global conditions are essentially related to the choice of
global transverse space. Not all local conditions are possible for a
given choice of transverse space. For instance, there is no
holomorphic function for a single $(d-3)$-brane in the Riemann
sphere\footnote{of course, one meets the same situation for other
  branes. However, for smaller branes one can always find harmonic
  functions with a single pole (describing a single brane) that lead
  to spaces asymptotically flat in transverse directions. This is not
  true for higher ($(d-3)$- and $(d-2)$-) branes).}.

To clarify these issues, let us consider the simplest solution in this
class: let us couple the action Eq.~(\ref{eq:SL2action}) to a charged
$(d-3)$-brane source. We first have to dualize the pseudoscalar $a$
into a $(d-2)$-form potential $A_{(d-2)}$ with field strength
$F_{(d-1)}=(d-1)\partial A_{(d-2)}$: $\partial a = e^{-2\varphi}\
{}^{\star}F_{(d-1)}$. The bulk plus brane action is

\begin{equation}
\begin{array}{rcl}
S & = & {\textstyle\frac{1}{16\pi G_{N}^{(d)}}}
{\displaystyle\int} d^{d}x \sqrt{|g|}\, \left\{R +{\textstyle\frac{1}{2}}
(\partial\varphi)^{2} +{\textstyle\frac{1}{2\cdot (d-1)!}}F_{(d-1)}^{2}
\right\} \\
& & \\
& & 
-\frac{T}{2}{\displaystyle\int} d^{d-2}\xi \sqrt{|\gamma|}
\left\{e^{\frac{2}{(d-2)}\varphi}\gamma^{ij}g_{ij} -(d-4) \right\}\\
& & \\
& & 
-\alpha\frac{T}{(d-2)!}{\displaystyle\int}
 d^{d-2}\xi A_{(d-2)\ i_{1}\cdots i_{(d-2)}}
\epsilon^{i_{1}\cdots i_{(d-2)}}\, ,\\
\end{array}
\end{equation}

\noindent where $g_{ij}$ and $A_{(d-2)\ i_{1}\cdots i_{(d-2)}}$ are the 
pullbacks through the embedding coordinates $X^{\mu}(\xi)$ of the
metric and $(d-2)$-form potential. $T$ is the tension (in principle, a
positive number) and $\alpha=\pm 1$ gives the sign of the charge
(which is evidently proportional to the tension). The coupling to
$\varphi$ is the only one that allows for solutions of the form we
want.

A solution is provided by

\begin{equation}
\left\{ 
\begin{array}{rcl}
ds^{2} & = & dt^{2}-d\vec{y}_{(d-3)}^{\ 2} -Hd\vec{x}_{2}^{\ 2}\, ,\\
& & \\
e^{-\varphi} & = & H\, ,\\
& & \\
A_{(d-2)\ t y^{1}\cdots y^{(d-3)}} & = & \alpha H^{-1}\, ,\\
& & \\
Y^{i} & = & \xi^{i}\, ,\hspace{1cm} \vec{X}_{2}=0\, ,\\
\end{array}
\right.
\end{equation}

\noindent where $H$ satisfies the equation

\begin{equation}
\partial^{2}H = -16\pi G_{N}^{(d)} T \delta^{(2)}(\vec{x}_{2})\, ,  
\end{equation}

\noindent i.e.~it is a harmonic function with a pole at $\vec{x}_{2}=0$,
where the brane is placed. The above equation is solved by a function
$H$ that behaves near $\vec{x}_{2}=0$

\begin{equation}
H\sim -8 G_{N}^{(d)}T \log{|\vec{x}_{2}|}\, .  
\end{equation}

\noindent It is clear that this solution cannot be globally correct as $H$
becomes negative for $|\vec{x}_{2}|>1$, but the local behavior of the
global solution has to be the same. Any solution behaving in this way
at any given point will describe a $(d-3)$-brane placed there. 

Let us now compute the charge. This is defined by

\begin{equation}
\label{eq:D7-branecharge}
p= \oint_{\gamma} e^{-2\varphi}\ {}^{\star}F_{(d-1)} =  
\oint_{\gamma} d a\, ,
\end{equation}

\noindent where $\gamma$ is a closed loop around the origin. 
$a$ is given by

\begin{equation}
\partial_{\underline{n}}a =\alpha \epsilon_{nm} \partial_{\underline{m}}H\, ,
\end{equation}

\noindent i.e.~combining $x^{1}+ix^{2}\equiv \omega$ 

\begin{equation}
\left\{
\begin{array}{rcl}
\partial_{\bar{\omega}} \tau & = & 0\, ,\hspace{1cm} \alpha =+1\, ,\\
& & \\
\partial_{\bar{\omega}} \bar{\tau} & = & 0\, ,\hspace{1cm} \alpha =-1\, ,\\
\end{array}
\right.  
\end{equation}

\noindent that is: $a$ is the real part of a holomorphic or 
antiholomorphic function of $\omega$, whose imaginary part is the
above function $H$. We find $a=\alpha 8 G_{N}^{(d)} T\, {\cal A}{\rm
  rg} (\omega)$ and $p=\alpha \frac{1}{16\pi G_{N}^{(d)}} T$. The
choice $\alpha=+1$ then, corresponds to a single $(d-3)$-brane with
charge $p=+\frac{1}{16\pi G_{N}^{(d)}} T$ placed at the origin and
corresponds to a holomorphic function $\tau ={\cal H}(\omega)$ that
close to the origin is given by

\begin{equation}
{\cal H} \sim -8 G_{N}^{(d)}T i \log{\omega}\, .   
\end{equation}

Observe that the charge is given by the multivaluedness of $\tau$
around the source, which goes from $\tau$ to $\tau +16\pi
G_{N}^{(d)}T$ which should be identified with $\tau$. The charge is
usually quantized due to quantum-mechanical reasons in multiples of
the unit of charge ($e$, say) which implies the identification
$\tau\equiv \tau+n e$ and the breaking of $SL(2,\mathbb{R})$. If $e=1$
(i.e.~$16\pi G_{N}^{(d)}T=1$ which we can always get by rescaling
$\tau$) then $SL(2,\mathbb{Z})$ is the unbroken symmetry of the theory
and the above $(d-3)$-branes are associated to the modular group
element $T$\footnote{For 10-dimensional type~IIB D7-branes $16\pi
  G_{N}^{(10)}=(2\pi)^{7}\ell_{s}^{8} g^{2}$ and $T=
  (2\pi)^{-7}\ell_{s}^{-8} g$, and, thus, ${\cal H} \sim
  -\frac{gi}{2\pi} \log{\omega}$. On the other hand,
  $C^{(8)}_{ty^{1}\cdots y^{7}}=g^{-1} H^{-1}$ ($\alpha=+1$) and we
  get $p=1$ in a most natural way.}.

We see that in this context solutions (and charges) can be
characterized by the non-trivial monodromies around singular points
which, by hypothesis, are elements of the modular group.

We can clearly generate via modular (duality) transformations of this
solution with $T$ monodromy other solutions with different
monodromies.  it is easy to see that if we perform a transformation
$\tau\rightarrow M(\tau)$ $M\in PSL(2\mathbb{Z})$ on the above
solution, the monodromy of the new solution around the origin will be
$MTM^{-1}$. The most interesting modular transformation is
$S(\tau)=-1/\tau$ which in other contexts relates electric and magnetic
(``S~dual'') objects. Then, the S~dual of the above solution will have
monodromy $STS$ around the origin and will be given either by ${\cal
  H}=-\frac{2\pi i}{\log{\omega}}$ using the general solution in the
form of Eq.~(\ref{eq:generalsolution1}) or with ${\cal
  H}=\frac{1}{2\pi i}\log{\omega}$ and the form (\ref{eq:otherform})
of the solution. This is the form we have used in the main text to
stress that we are dealing with a solution different from the one with
monodromy $T$, the difference being in the choice of holomorphic
function since, as we have stressed at the beginning of this Appendix
all homomorphic solutions can always be written in the form
(\ref{eq:generalsolution1}), no matter if the monodromy is $T$ or
$STS$.


\section{The KK Origin of the $SL(2,\mathbb{R})/SO(2)$ Model}
\label{sec-sl2z}

We are going to see how the modular group $PSL(2,\mathbb{Z})
=SL(2,\mathbb{Z})/\{\pm\mathbb{I}_{2\times 2}\}$ and the
$SL(2,\mathbb{Z})/SO(2)$ sigma model arise in standard Kaluza-Klein
compactification on a 2-torus $T^{2}$.


\subsection{The Modular Group}

As usual in KK compactifications, we use two periodic coordinates
$x^{m}$ $m=1,2$ whose periodicity is fixed to $2\pi\ell$ where $\ell$
is some fundamental length. This means that we make the
identifications

\begin{equation}
\vec{x}\sim\vec{x}+2\pi\ell \vec{n}\, ,
\hspace{1cm}
\vec{x}=
\left(
  \begin{array}{c}
x^{1}\\ x^{2}\\
  \end{array}
\right)\, , 
\hspace{1cm}
\vec{n}\in\mathbb{Z}^{2}\, . 
\end{equation}

The information on relative sizes and angles of the periods and the
size of the torus is codified in the internal metric $G_{mn}$, 

\begin{equation}
ds^{2}_{\rm Int}=d\vec{x}^{\, T}Gd\vec{x}\, ,  
\end{equation}

\noindent which is, by hypothesis, independent on the torus 
coordinates $\vec{x}$, (but may depend on the remaining coordinates).

The KK Ansatz is invariant under global diffeomorphisms in the
internal manifold. These are, generically, of the form

\begin{equation}
\vec{x}^{\prime}=R^{-1}\vec{x}+\vec{a}\, ,
\hspace{1cm}
R\in GL(2,\mathbb{R})  
\vec{a}\in \mathbb{R}^{2}\, .
\end{equation}

$\vec{a}$ simply shifts the coordinate origin and does not affect the metric.
$R$ acts on the internal metric according to

\begin{equation}
G^{\prime}=R^{T}GR\, ,
\hspace{1cm}
(G_{mn}=R^{p}{}_{m}G_{pq}R^{q}{}_{n})\, .  
\end{equation}

We want to separate the volume part of the metric from the
rest\footnote{This is necessary, for instance, when we are interested
  in conformal classes of equivalence of metrics, as in string path
  integrals, but convenient in general.}. Thus, we
define\footnote{Remember that $G$ has signature $(--)$.}

\begin{equation}
K\equiv |{\rm det} G_{mn}|\, ,
\hspace{1cm}
G_{mn}\equiv -K^{1/2}{\cal M}_{mn}\, .
\end{equation}

\noindent ${\cal M}$ has determinant $+1$ and, therefore, it is a symmetric
$SL(2,\mathbb{R})$ matrix and, in fact, it can be understood as an
element of the coset $SL(2,\mathbb{R})/SO(2)$ with only two
independent entries. If we factor out the determinant of the
$GL(2,\mathbb{R})$ transformations too,

\begin{equation}
R\equiv |{\rm det} R^{m}{}_{n}|\, ,
\hspace{1cm}
s={\rm sign}({\rm det} R^{m}{}_{n})\, ,
\hspace{1cm}
R^{m}{}_{n}\equiv sR^{1/2}{\cal S}^{m}{}_{n}\, ,
\end{equation}

\noindent  then the volume element $K$ and the matrix ${\cal M}$
transform according to

\begin{equation}
\label{eq:Mtransformationrule}
\begin{array}{rcl}
{\cal M}^{\prime} & = & S^{T}{\cal M}S\, ,\\
& & \\
K^{\prime} & = & RK\, .\\
\end{array}
\end{equation}

$|K|$ is an element of the multiplicative group $\mathbb{R}^{+}$ and
$S$ is an element of $SL(2,\mathbb{R})$. This decomposition reflects
the decomposition $GL(2,\mathbb{R}) =SL(2,\mathbb{R})\times
\mathbb{R}^{+} \times \mathbb{Z}_{2}$. $s$ does not act neither on
$K$ nor on ${\cal M}$.

We have not yet taken into account the periodic boundary conditions of
the coordinates, that have to be preserved by the diffeomorphisms in
the KK setting. Clearly the rescalings $R$ do not respect the torus
boundary conditions, but they rescale $\ell$. The rotations $S$
respect the boundary conditions only if $S^{-1}\vec{n}\in
\mathbb{Z}^{2}$ the matrix entries are integer, i.e.~$S\in
SL(2,\mathbb{Z})$. Up to a reflection $S=-\mathbb{I}_{2\times 2}$,
these diffeomorphisms are known as {\it Dehn twists} and are not
connected with the identity (in fact, they constitute the mapping
class group of torus diffeomorphisms) and they constitute the {\it
  modular group} $PSL(2,\mathbb{Z})
=SL(2,\mathbb{Z})/\{\pm\mathbb{I}_{2\times 2}\}$. This is the group
that acts on ${\cal M}$.

We are going to write the modular group matrices in the slightly
unconventional form

\begin{equation}
\label{eq:modtramatrix}
S= 
\left(
  \begin{array}{cc}
\alpha & \,\,\,\gamma \\
& \\
\beta & \delta \\
  \end{array}
\right)\, ,
\end{equation}

\noindent to get the conventional form of the transformation of the 
modular parameter Eq.~(\ref{eq:modtrans}).


\subsection{The Modular Parameter $\tau$}

We can define a complex modular-invariant coordinate $\omega$ on $T^{2}$
by 

\begin{equation}
\omega ={\textstyle\frac{1}{2\pi\ell}}\, \vec{\omega}^{T}\cdot \vec{x}\, ,
\hspace{1cm}
\vec{\omega}=\mathbb{C}^{2}\, ,  
\end{equation}

\noindent where, under modular transformations, we assume that the 
complex vector $\vec{\omega}$ transforms according to

\begin{equation}
\label{eq:compelxperiodstransformationrule}
\vec{\omega}^{\prime} = S^{T}\vec{\omega}\, .
\end{equation}

\noindent The periodicity of $\omega$ is

\begin{equation}
\omega \sim \omega +\vec{\omega}^{\, T}\cdot \vec{n}\, ,
\hspace{1cm} \vec{n}\in \mathbb{Z}^{2}\, .  
\end{equation}

What we have done is to transfer the information contained in the
metric (more precisely, in ${\cal M}$) into the complex periods
$\vec{\omega}$.  The relation between these two is

\begin{equation}
{\cal M} =
\frac{1}{\Im{\rm m}(\omega_{1}\bar{\omega_{2}})}
\left(
\begin{array}{cc}
|\omega_{1}|^{2} & \Re{\rm e}(\omega_{1}\bar{\omega_{2}})\\
& \\
\Re{\rm e}(\omega_{1}\bar{\omega_{2}}) & |\omega_{2}|^{2} \\
\end{array}
\right)\, .  
\end{equation}

We can check that the transformation rules for the complex periods
Eq.~(\ref{eq:compelxperiodstransformationrule}) and for the matrix
${\cal M}$ Eq.~(\ref{eq:Mtransformationrule}) are perfectly
compatible.

In terms of the modular-invariant complex coordinate, the torus metric
element takes the form

\begin{equation}
ds^{2}_{\rm Int}=K^{1/2}\frac{1}{\Im{\rm m}\omega_{1}\bar{\omega_{2}}}
d\omega d\bar{\omega}\, .  
\end{equation}

Observe that $\Im{\rm m}(\omega_{1}\bar{\omega_{2}})$ is
modular-invariant (and a quite important one).

It should be clear that not all pairs of complex periods characterize
different tori. Recall that ${\cal M}$ only has 2 independent entries
while $\vec{\omega}$ contains 4 real independent quantities. In
particular, we can see that multiplying $\vec{\omega}$ by any complex
number leaves the matrix ${\cal M}$ invariant. It is customary to
multiply by $\omega_{2}^{-1}$ both the coordinate $\omega$ and define

\begin{equation}
\xi=\omega/\omega_{2}\, ,
\hspace{1cm}
\tau=\omega_{1}/\omega_{2}\, ,  
\end{equation}

\noindent that can always be chosen to belong to the upper half complex 
plane $\mathbb{H}$ $\Im{\rm m}(\tau)\geq 0$ ($-\omega_{1}$ defines the
same torus as $\omega_{1}$).

Under a modular transformation with $S$ given by
Eq.~(\ref{eq:modtramatrix}), the modular parameter undergoes a
fractional-linear transformation

\begin{equation}
\label{eq:modtrans}
\tau^{\prime} = \frac{\alpha\tau +\beta}{\gamma\tau +\delta}\, .  
\end{equation}

\noindent and the torus coordinate $\xi$ transforms

\begin{equation}
\xi^{\prime} =\frac{\xi}{(c\tau +d)}\, .  
\end{equation}

\noindent Finally, in terms of $\tau$, the matrix ${\cal M}$ reads

\begin{equation}
\label{eq:Mparametrizedbytau}
{\cal M} =
\frac{1}{\Im{\rm m}(\tau)}
\left(
\begin{array}{cc}
|\tau|^{2} & \Re{\rm e}(\tau)\\
& \\
\Re{\rm e}(\tau) & 1 \\
\end{array}
\right)\, .  
\end{equation}


\subsection{The $SL(2,\mathbb{R})/SO(2)$ Sigma-Model}

In pure KK theory (with no higher-dimensional fields apart from the
metric), the toroidal compactification of the Einstein-Hilbert action
from $\hat{d}$ to $d$ dimensions with the KK Ansatz

\begin{equation}
\left( \hat{e}_{\hat{\mu}}{}^{\hat{a}} \right) = 
\left(
\begin{array}{cr}
e_{\mu}{}^{a} & e_{m}{}^{i}A^{m}{}_{\mu} \\
&\\
0             & e_{m}{}^{i}                \\
\end{array}
\right)\, , 
\end{equation}

\noindent where the internal metric

\begin{equation}
G_{mn}=e_{m}{}^{i}e_{n\, j}=-e_{m}{}^{i}e_{n}{}^{j}\delta_{ij}\, .
\end{equation}

\noindent gives, upon the rescaling

\begin{equation}
g_{E\, \mu\nu}= K^{\frac{2}{(d-2)}} g_{\mu\nu}\, ,  
\end{equation}

\begin{equation}
  \begin{array}{rcl}
S & = & {\displaystyle\int} d^{\hat{d}}\hat{x}\sqrt{|\hat{g}|}\, \hat{R}\\
& & \\
& = & 
{\displaystyle\int} d^{d}x \sqrt{|g_{E}|}\left[R_{E}
+{\textstyle\frac{(\hat{d}-2)(\hat{d}-d)}{4(d-2)}}
(\partial \log{K})^{2}
+{\textstyle\frac{1}{4}}{\rm Tr}\left(\partial{\cal M}{\cal M}^{-1}\right)^{2}
\right.\\
& & \\
& &
\left.
\hspace{2cm}-{\textstyle\frac{1}{4}} K^{\frac{(\hat{d}-2)}{(d-2)}}
{\cal M}_{mn}F^{m\, \mu\nu}F^{n}{}_{\mu\nu}\right] \, .\\
\end{array}
\end{equation}

\noindent The kinetic term for the scalar matrix ${\cal M}$ is manifestly 
invariant under $SL(2,\mathbb{R})$ transformations (the action we
started from is diffeomorphism-invariant). Using the parametrization
Eq.~(\ref{eq:Mparametrizedbytau}), it takes the standard form

\begin{equation}
{\textstyle\frac{1}{2}} 
\frac{\partial\tau\partial\bar{\tau}}{(\Im{\rm m}(\tau))^{2}}\, .
\end{equation}


\end{document}